# APPLYING USER EXPERIENCE AND USER-CENTERED DESIGN SOFTWARE PROCESSES IN UNDERGRADUATE MOBILE APPLICATION DEVELOPMENT TEACHING

Manuel Ignacio Castillo López[1], Ana Libia Eslava Cervantes[2] and Gustavo de la Cruz Martínez[2]

[1]Facultad de Ciencias, Universidad Nacional Autónoma de México, Mexico City, Mexico – ORCID: 0000-0002-2307-5860
[2]Instituto de Ciencias Aplicadas y Tecnología, Universidad Nacional Autónoma de México, Mexico City, Mexico

## ABSTRACT

*Agile methods in undergraduate courses have been explored by various authors looking to close the gap between industry and professional profiles. We have structured an Android application development course based on a tailored agile process for development of educational software tools. This process is based on both Scrum and Extreme Programming in combination with User Experience (UX) and User-Centered Design (UCD) approaches. The course is executed in two phases: the first half of the course's semester presents theory on agile and mobile applications development, the latter half is managed as a workshop where students develop for an actual client. The introduction of UX and UCD exploiting the close relationship with stakeholders expected from an agile process can enhance Quality in Use features. Since 2019 two of the projects have been extended in agreement between the client and students. Students, clients and users have found value in the generated products.*

## 1. INTRODUCTION

In recent years, the *grupo Espacios y Sistemas Interactivos para la Educación* (ESIE) from the *Instituto de Ciencias Aplicadas y Tecnología* (ICAT) of the *Universidad Nacional Autónoma de México* (UNAM), has worked on improving their software development processes by including in them User Experiences (UX) design practices [1]. Moreover, also in recent years various authors have explored the inclusion of agile processes in software development teaching [2] – [4].

The motives to include UCD in software development processes are related to Human-Factors software quality attributes, which have been identified as primary factors on user retention and success of a software project [5] – [7].

The grupo ESIE works on tools and spaces development for teaching or didactic activities, for this purpose, they include gamification practices for *interactive experiences* building. The goal of the products from grupo ESIE projects is to convey these experiences to users in order to guide them through the learning or reinforcement of knowledge activities. Specifically, the goal of the grupo ESIE in including UX design practices into its software process, is to dispose the adequate tools for product design [1].

The way in which the group has approached the inclusion of Human-Factors in its software processes, leans on the process exercise in courses its members taught as part of their academic activities at UNAM. In particular, the process of the grupo ESIE has been applied repeatedly in the course *Programación de Dispositivos Móviles* (PDM) of the *Licenciatura en Ciencias de la Computación* (LCC) at the *Facultad de Ciencias* (FC), UNAM; a Bachelor degree program on Computer Science. The application of the process introduces students to UCD and UX practices, while grupo ESIE members observe the effects of the process activities in the students' course projects.

In this document, a reflection of the results obtained so far by applying the grupo ESIE process in the PDM course is presented, considering the means in which Human-Factors from an UX perspective change the priority of requirement analysis and design activities and the impact the process has on students projects.

The document is organized as follows: in the next section we present similar approaches where agile methodologies have been applied into undergraduate courses, then we present a general description of the grupo ESIE process, chapter four describes the PDM course and chapter five shows how the grupo ESIE process is applied. Later a discussion on the results obtained after several PDM course instances have occurred is presented and then we finalize by reflecting about course project's stakeholder satisfaction.

## 2. AGILE PROCESSES INCLUSION IN UNDERGRADUATE TEACHING

In recent years several proposals have emerged which explore the effects of incorporating agile processes in undergraduate courses. A recent work in this line is from Lara & Figueroa [2] who apply *Extreme Programming* (XP) to a Software Engineering course which objective is the production of educational applications for mobile devices. The course is divided in phases implied by XP in combination with teaching objectives: teaching requirements approach, pedagogical content design, user interface design, implementation, testing and delivery.

Their requirement acquisition strategy is based on User Histories, from which they define tasks to conform the project's iterations. The solution design employs class diagrams, prototypes and navigation maps. Tests consists on automated unit testing. From the release of their paper, the proposed course has begun offering the development of teaching applications to Argentine educational institutions, but it only had gotten to the requirements acquisition phase [2].

Tesei et. al [3] consider necessary to incorporate agile process into teaching process to close the gap between professional profiles and industry needs. For this purpose, they have designed a security management software project to be developed in a course about software development. The students are given development roles while teachers act as Product owners, some students are chosen as Scrum Masters which won't perform coding activities directly. The course spans 16 weekly sessions divided in 4 week-iterations. The first iteration is focused on work environment setup and solution design. Each session has a 4 hour length, the first two hours are dedicated to explore theory about agile methodologies, the next two hours are used to apply such theory; Scrum ceremonies are performed in each session [3].

As in the work by Lara & Figueroa [2], design is based on User Histories. Students are allowed to renegotiate with the Product owner the scope and deliverables for each iteration. At the time their paper was published the course had not finished yet, but the students where surveyed and they pointed out that the learning of agile methodologies has been clear and useful through the project [3].

The work of Martín Gómez [4] is similar to that of Tesei et. al [3] as the goal is to apply agile methodologies to encourage group work and digital competences, closing the gap between professional profiles and industry needs. The author points that the educational system shares the primacy of people and interpersonal relationships with agile methodologies so in combination they should allow the development of autonomy, capabilities and abilities.

Their proposal is to implement Scrum in combination with Kanban in a group of six students with the following objectives: to develop autonomous learning competences, develop social abilities such as communication, leadership and technological competences. Another similarity with the work of Tesei et. al [3] is the distribution of Scrum roles: the professor acts as Product owner and most of students acts as developers with a Scrum Master in the team [4].

## 3. GRUPO ESIE'S SOFTWARE PROCESS

The grupo ESIE's software process has been developed from the perspective of the UCD, leaning in UX design practices [1]. The process is divided in six phases that occur iteratively according to agile strategies and principles. Its application in projects can be described through 3 stages which represent the main objectives of the phases each stage contains.

The six phases are described next, grouped in their correspondent stage [1]:

### 3.1. First stage. Preproduction

This stage contains the *Definition* and *Design* phases. In this stage the project is prepared: its objectives, scope and requirements are established, and ends with the deployment of a high-fidelity prototype, a *minimum viable product* which will evolve in the next stages until the final product is obtained.

*Phase 1) Definition*. Initial agreements are established with clients and stakeholders, their needs are documented and the project objectives are defined.
*Phase 2) Design*. The product's central concept is defined and the specification of experiences and characteristics of the product are described.

### 3.2. Second stage. Production

Contains the *Realization* phase and shares the *Testing and validation* phase with the next stage. Tasks are selected to be developed in a short period of time by applying agile methodologies. These tasks will be performed on top of the results of the last iteration. Results can be assessed to obtain feedback for the next iteration, this allows timely identification of deficiencies on the design or its implementation so they can be improved.

*Phase 3) Realization*. The design obtained in the last stage is implemented on the results of last iteration.
*Phase 4) Testing and validation.* The product is tested with target users employing usability testing techniques in a controlled environment.

### 3.3. Third stage. Postproduction

Contains the C*losing* and *Feedback and maintenance* phases. The generated product is validated according to the specification agreed with the client. If the product's quality is determined to be within expectations, results are presented to the client who will decide if the product is ready to be Deployed or if a new *Production* stage is required (this also depends on the scope and agreements).

*Phase 5) Closing*. The product is Deployed along with any other deliverables agreed with the client.
*Phase 6) Feedback and maintenance*. After the product is deployed, it is studied if the agreements with the client allows it. Modifications to enhance the generated experience can be considered for a next version of the project.

## 4. APPLICATION OF THE GRUPO ESIE'S PROCESS IN THE PDM COURSE

The activities of the grupo ESIE members include teaching. The PDM course is a regularly taught by some of the group members. This course has a theoretical-practical approach, concepts around mobile applications development are presented during the first half of the semester. These concepts include general mobile applications interface design, development with Android Studio and the Android SDK. Student grading is mainly done by assessing the software they develop by assignments and a final project with an actual client.

Figure 1 shows students practicing Scrum though a low-fidelity prototype. Figure 2 shows students practicing Kanban for task management
.

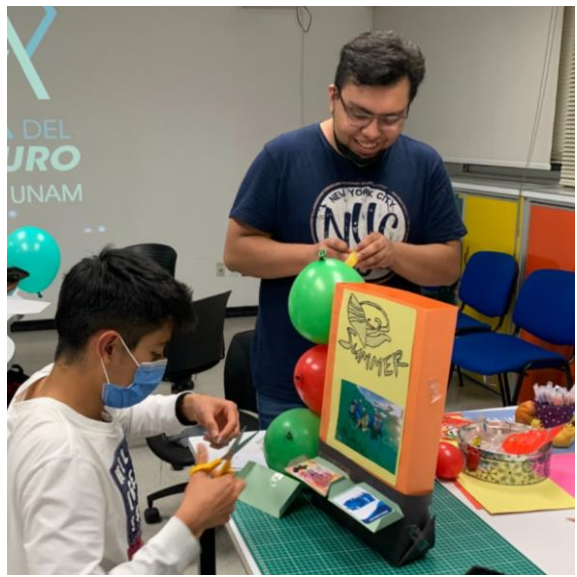

Figure 1. Students practice the Scrum methodology through a low-fidelity prototype

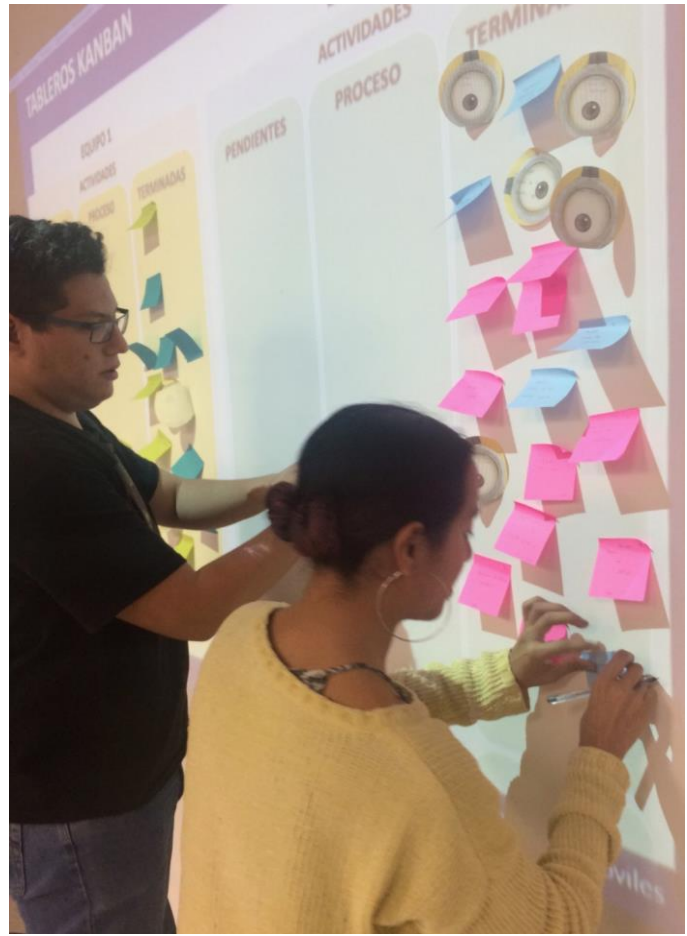

Figure 2. Students practice project task management through a Kanban board

The second half of the course focuses on development of a software project with a client who usually is another UNAM academic who has a need around a mobile application. The project's objective is that the students build a minimum viable product of the application specified by the client. The project applies the grupo ESIE's process, which has allowed multiple students groups to develop such applications, satisfying the client and users' highest priority requirements thought UX and UCD activities in the process.

The students developed products are assessed with target users at the ICAT's *Aula del Futuro*, which is an specialized environment used for UX and usability testing with a focus on learning spaces and tools. Results from user testing is shared with the student's client, which are measured using *System Usability Scale* (SUS) [8]. SUS estimates the degree with which the product allows users to reach their goals under the expected use-context [9], [5], [6]. Figure 3 shows students performing a usability test at the *Aula del Futuro*.

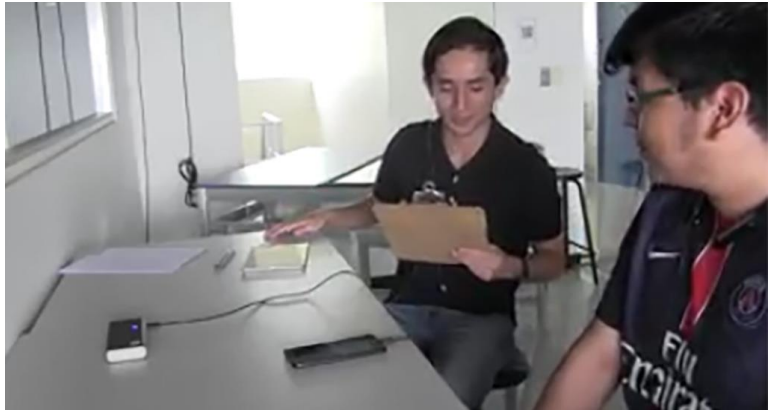

Figure 3. A student guides a user through a usability test

Project development is done by teams of 3 to 5 students, where one of the team members has a *Facilitator* role, who is responsible of inspect the task's progress though a Kanban board. Students are instructed to only change the tasks they are directly responsible for. The *Facilitator* also moderates the interaction between team members and assist them to get the necessary work resources. He usually manages the team's software repositories.

The other team members choose a development role, such as Database Administrator or Designer; every team member including the *Facilitator* must perform programming tasks.

FC courses have two teachers: a professor who is administratively responsible for the course and one or more professor's assistants. The professor has a Coach role within the students projects, providing assistance, orientation and feedback about their process implementation and results. Professor's assistants have a Product owner role, reviewing the product building, orienting and providing feedback about the software artifacts the students generate.

## 4.1. Preproduction stage

Following the 3 stages in which the six phases of the grupo ESIE's process, we begin the description of the students activities by the *Preproduction* stage in the PDM course, starting in the *Definition* phase where students must prepare an interview with the client for requirements acquisition, client needs and expectations are studied. Then the *Design* phase begins as the students prepare a first product proposal to be presented to the client and further refine their design.

Student's proposal include a target user profile analysis for UCD tasks development through the Persona technique [10]. The user profile is obtained from questionnaires published online, in forums and other web sites where target users are expected to be found. The proposal is built around the user's profile, client requirements and expectations. In particular students are requested to focus on navigation and user interface design. The proposal is prepared in a low-fidelity prototype usually built on paper and which represents the most significative functionalities. From this first prototype the proposal is further refined and a second prototype is produced, this second prototype is then presented to the client along the full proposal.

### 4.2. Production stage

The *Realization* phase starts with the observations made by the client about the student's proposal for a next refining and task prioritizing round. The release cycle has a length between one and two months with iterations of one or two weeks long.

The prioritization of tasks is reflected on a Kanban board and an Android Studio project is prepared in a team controlled Git-based repository. Though *Continuous Integration* and *Continuous Delivery*, students carry their programming tasks in individual git-branches and at the end of every iteration changes are integrated into a branch which represents the final build.

### 4.3. Postproduction stage

When the *Production* of the application is done, the generated product is *Tested and Validated*. The product should show the viability of the proposal made by the students. To appraise the degree with which the product answers to the client and user's needs, usability tests are performed observing user interaction sessions in a usability laboratory. For these tests, students must develop a series of assessment instruments and a test-session script which dictates the activities to perform at the test, particularly those activities which the user must perform through the product which should match with the most relevant functionalities identified. An exit questionary is also developed to collect the user's opinion on the effectiveness of the product to allow them to perform the requested tasks, these questionnaires are structured to be evaluated with the SUS scale. Observations of the tests and answers to assessment instruments and questionnaires are analyzed to produce recommendations which should be addressed in a next version.

The application of the grupo ESIE's process in the PDM course ends in the *Closing* phase. Tests results are presented to the user along with the tested product. The *Feedback and maintenance* can happen if new agreements are made between the client, grupo ESIE members and the students for a new version of the project, which would be part of the student's graduation activities.

## 5. GRUPO ESIE'S PROCESS ACTIVITIES IN STUDENT'S TASKS

Grupo ESIE's process defines some general activities to be performed along a software project, with specific inputs and required conditions for the described activities as well their specific outputs. These process' activities are derived from Scrum and Extreme Programming practices. Applying the grupo ESIE's process in the PDM course's projects offers a familiar environment for the group members who coach the students developing mobile applications and encompasses task planning and distribution, quality frameworks and UCD practices for the students to develop their applications on.

During the project, at the beginning of each process's phase, it is presented to the students the work strategies for the phase's activities, their inputs, outputs and required conditions. The process's activities are integrated into the PDM course as student's tasks and their outputs are graded. Students plan their tasks according to the current phase's framework and then they practice the process's activities.

By the course's end the students recognize various of the agile principles in which the grupo ESIE's process are based on and the effects these principles have on their work. Students who continue to develop the applications have followed the learned practices and built software artifacts with a UCD approach.

## 5.1. UX and DCU application in the PDM course

It has been repeatedly observed that UCD and UX activities are foreign to the students at the beginning of the course. When the first tasks at the *Design* phase occur, it is likely for the students to focus on functional requirements, while quality in use characteristics, user and client expectations are initially dismissed. As tasks develop and the identification of usability traits become more apparent to the students, they show a change of attitude towards UCD. The value of usability testing as a mean for user quality assessment is usually highly regarded by students who finish the course.

## 5.2. Agile methodology application

Grupo ESIE's process is based on agile methodologies, with emphasis in collaboration principles with the client, users and other stakeholders who are seen as part of the development team. Agile methodologies offer a framework with conditions and mechanisms which enable collaboration and feedback required by user-centered approaches [5], [6], [11].

Along with UCD and UX activities, it has been frequently observed that task management, distribution and software repository management activities, such as versioning, become clearer by the end of the semester, even when students have had previous software product development experiences.

This may be the effect of the grading scheme, which not only grades the product, documentation, artifacts and development history in the project's software repository, but also includes task distribution and timeliness. Task management is graded though activity and history of the project's kanban board at the end of each sprint. After reviewing each team's kanban board, feedback is provided to the members of the team about their task management, as the project evolves, students refine their understanding of agile concepts, challenges and risks around team management.

## 5.3. Application of usability testing

The implementation of usability testing with target users at a usability laboratory shows students how users react to both defects and features they like and the impact such reaction has on their desire to keep using or to reject the product. Product assessment allows students to measure their solution's usability and show the client these results so conformance with his needs is further exhibit beyond the required functionality.

## 6. PROJECTS CONTINUATED BEYOND THE PDM COURSE

The PDM course has been taught yearly using the described process since 2019. Two of the course projects have been continued after the semester in which these projects occurred ended. The students participate in the new version project as part of their graduation process.

The first project which had continuity is *Hocus Focus*, an application designed for people with Attention Deficit – Hyperactivity Disorder (ADHD). The application offers note taking functionality and mainly an habit agenda as show in figure 4 (left). Habits are associated with a daytime and an alarm (see right of figure 4) so ADHD users can keep a constant rhythm with their activities by following the alarms. *Hocus Focus* was awarded the *Premio a la Innovación UNAM 2019*, which is an annual prize given every year since 2018 to recognize innovative research and entrepreneurial projects and ideas. *Hocus Focus* was awarded the second prize in the social innovation category.

Another continued project from the PDM course was *Polyx*, a polymers teaching-tool application for a course in a chemistry degree program theached at UNAM's Facultad de Química (Chemistry Faculty). The application contains theoretical concepts around polymers as show in figure 5 (letf), and gamified self-reviews about the concepts (see right of figure 5). Other UNAM's dependencies have shown interest in the application of the content/self-rewiew structure of *Polyx* in different contexts.

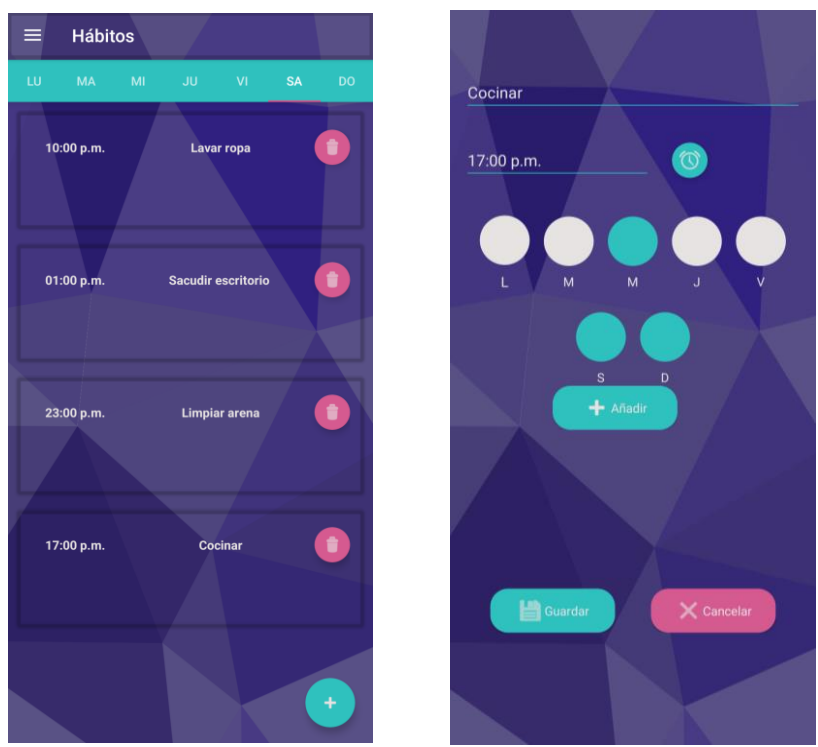

Figure 4. Some user interfaces of the *Hocus Focus* application. Left shows the *habit agenda* where user's habits are grouped by weekly occurrence. Right shows the habit registry where habits are setup for weekly occurrence at an specific time and a custom alarm ringtone.

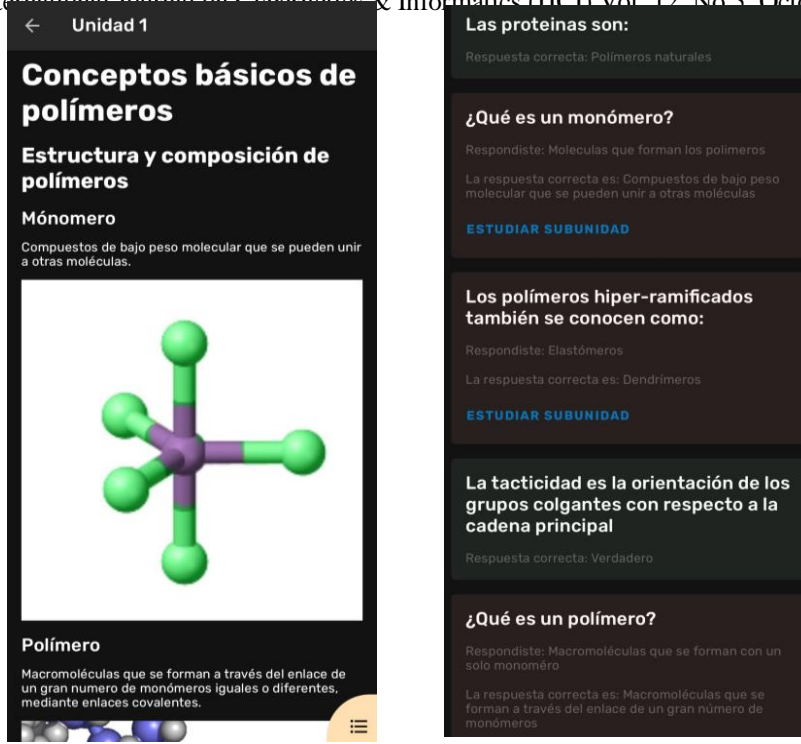

Figure 5. Some user interfaces of Polyx. Left shows the theoretical content organized topically. Right shows the result of a self-review test where it details which answers were right and which were wrong, giving the right answer and linking the theory to it**.**

## 7. CONCLUSIONS

Similarly to the related work presented in section 2, the methodology applied in the PDM course combines practices from both Scrum and Extreme Programming. It has applied the grupo ESIE's methodology annually since 2019 in collaboration with academics external to the group to develop minimal viable products for diverse domains.

Student who have taken the PDM course and later incorporated into the group's projects, have shown an ease to structure and manage their activities and a reflective attitude towards the project goals. More over, the development experience from the course helps them to grasp a clearer vision of the activities they perform within the project, including their requirements, expected results and delivereables. Collectively, this attitude enables a fluent and reflective evolution of the projects.

This also has provided feedback to further refine the process, identifying ambiguous activities in it. On the other hand, the application of the process has helped students to practice UCD and to build Android applications which satisfy their requirements, which should match the most relevant characteristics expected by the client and are presented to users through usable interfaces. Students also collect feedback from users to identify defects and opportunities that should be addressed in future versions.

Students also have shown the capacity to recreate UCD tasks in other projects and develop them by iterative and incremental means. On continuated projects, the resulting project have become more complete and more correct in respect to the client's goals. Quality in use aspects are carefully managed, any modification to the user interaction has to be designed, implemented and tested as specified by the grupo ESIE's process.

## 7.1. Application of grupo ESIE's process

Grupo ESIE's process is based on agile methodologies, UCD and UX design, its application on the PDM course preserves the basic concepts from all these software development approaches which results on prioritization of the preproduction stage. At first glance this appears to work against agile principles such as *working software over comprehensive documentation* and *responding to change over following a plan* [12].

Several agile projects show a literal interpretation of the agile manifesto as a software methodology, although the manifest is a high-level synthesis which describes a framework but lacks assessment tools and directives to properly define a methodology [11, 12, 13]. For example Tesei et. al [3] points out as an advantage of their proposal the reduction of documentation and the start of coding activities in early stages of the project.

While it is possible to build software with minimal documentation, limited documentation usually has negative impacts on software lifecycle stages after its lauch. Agile methodologies proposed or promoted by the agile manifesto authors have mechanisms for delay or automatize documentation production, to produce informal or non-conventional documentation which is as useful but easier to manage than traditional documentation; documentation is not ignored nor removed from an agile project [11, 14, 15].

PDM course's students who have used development strategies with minimal documentation management have found the preproduction stage as cumbersome, since the first weeks of the course's project are dedicated to design activities which usually involve documentation and low-fidelity prototypes production, but very few (if any) coding tasks. By the end of the course, most of these students understand the practicality of an exhaustive design stage since it offers better planning opportunities based on a well-structured design, and development begins from a prototype which can be evolved in the final product in about a month.

Furthermore, the produced design and documentation generated in the preproduction stages are essential for projects extensions after the PDM course's end of semester, when agreements are reached between the grupo ESIE, students and the projects' clients.

Moreover the application of the grupo ESIE's process in the PDM course requires autonomy from students, their proposals and designs from the clients' requirements are developed into several work products which are graded. This methodology is very similar to educational models which are being preferred instead of classical models of concept and process memorization [4]. Grading of several work products, such as Kanban board changes, software repository changes through version control, design documents, prototypes, code quality and usability performance, allow punctual and rich feedback which students can use to reflect on their working strategies and process performance.

### ACKNOWLEDGEMENTS

The authors would like to thank the academics and students who have make and effort for the development of projects derived from the PDM course, in particular we would like to thank Lupita Vásquez Fabela for being the client for *Hocus Focus*, application developed by the students: Adrián F. Vélez Rivera, Aide I. García Hernández, Jesús M. Colín Torres and Raúl

Ascencio Bolio. Also we would like to thank Yara C. Almanza Arjona for being the client for *Polyx*, developed by the students: Joshua J. Pedrero Gómez, Mauricio Araujo Chávez, Paola Vázquez Rizo and Uriel Rosas Franco.

# REFERENCES

[1] G. de la Cruz Martínez, A. L. Eslava Cervantes, and R. Castañeda Martínez, «Diseño de la Experiencia del Usuario para Espacios Interactivos de Aprendizaje no Formal», Res. Comput. Sci., vol. 89, pp. 53-62, 2015.

[2] C. Lara and L. Figueroa, «Metodología ágil para el desarrollo de aplicaciones móviles educativas» in L. Cecchi, S. Roger y A. Buccella (eds.), TE&ET 2020: Libro de Actas XV Congreso de Tecnología en Educación y Educación en Tecnología – REDUNCI, July 6 and 7 2020, Argentina, Universidad Nacional del Comahue, pp. 206-213

[3] F. Tesei, M. Cabrera, M. Vaquero, and D. Tedini, «Acercando la academia al mundo real: una experiencia de aplicación de Metodologías Agile al proceso de enseñanza-aprendizaje en una asignatura de desarrollo de software» in Sociedad Argentina de Informática e Investigación de Operaciones (coord.), I Simposio Argentino de Educación en Informática SAEI 2019, September 16 – 20 2019, Argentina, Universidad Nacional de Salta, pp. 100-116.

[4] S. Martín Gómez, «Aplicación de las Metodologías Ágiles al proceso de enseñanza-aprendizaje universitario», Revista d'Innovació Docent Universitària, no. 12, pp. 62-73, 2020.

[5] N. O. Bernsen and L. Dybkjaer, Multimodal usability. Springer, 2009.

[6] A. Dix, J. Finlay, G. D. Abowd, and R. Beale, Human-computer interaction, 3rd ed. Harlow, England; New York: Pearson/Prentice-Hall, 2004.

[7] P. Bourque, R. E. Fairley, and I. C. Society, Guide to the Software Engineering Body of Knowledge (SWEBOK(R)): Version 3.0, 3rd ed. Washington, DC, USA: IEEE Computer Society Press, 2014.

[8] J. Brooke, «SUS: A quick and dirty usability scale», in Usability Evaluation in Industry, W. Jordan, B. Thomas, B. A. Weerdmeester, y A. L. McClelland, Eds. London: Taylor and Francis, 1996.

[9] J. Brooke, «SUS: A retrospective», J. Usability Stud., vol. 8, no. 2, pp. 29-40, feb. 2013.

[10] A. Cooper, R. Reimann, and D. Cronin, About Face 3: The Essentials of Interaction Design. 2007.

[11] I. Sommerville, Software engineering. Boston: Addison-Wesley, 2011.

[12] K. Beck et al., «Manifesto for Agile Software Development», Manifesto for Agile Software Development, 2001. https://agilemanifesto.org/ (accedido 9 de febrero de 2023).

[13] J. Highsmith, «History: The Agile Manifesto», History: The Agile Manifesto, 2001. https://agilemanifesto.org/history.html (accedido 9 de febrero de 2023).

[14] R. C. Martin and M. Martin, Agile principles, patterns, and practices in C#. Upper Saddle River, NJ: Prentice Hall, 2007.

[15] R. C. Martin, Clean Code: A Handbook of Agile Software Craftsmanship, 1.ª ed. USA: Prentice Hall PTR, 2008.

# AUTHORS

**Manuel Ignacio Castillo López** Universidad Nacional Autónoma de México, Master in Computer Science. Professor Assistant (PA) at UNAM's FC and Sc.D student in the field of Software Process and Product Quality. Ignacio has collaborated with the grupo ESIE since 2016 in product development and process evolution. He also has been PA for the PDM course on several occasions since 2019. His interest in software quality comes from more than 10 years as a professional software developer and project manager, having encountered many of the challenges faced in the software industry and their hard outcomes, including layoffs and business cease of operations.

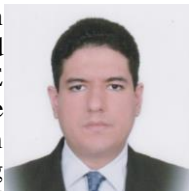

**Ana Libia Eslava Cervantes** Universidad Oberta de Catalunya, Master in Multimedia Desing. Member of the grupo ESIE, ICAT, UNAM since 2015 and co-founder of the Aula del Futuro project, which aims to create a replicable model of a computer assisted teaching space. Libia is also the PDM course's professor. The Aula del Futuro model not only offers a computar-based framework but also teaching practices to make an efficient use of digital tools. The model already has several implementations around Mexico and Latin America.

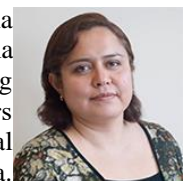

Her work focus is on pedagogical innovation; design, development and assessment of interactive teaching environments with an UCD approach and of teaching strategies.

**Gustavo de la Cruz Martínez** Universidad Nacional Autónoma de México, Sc.D in Computer Science. Founding member of the project of the Aula del Futuro. Professor at UNAM's FC since 2002 and full-time Academic at ICAT, UNAM since 2005. His areas of interest are human-computer interaction, user modeling and user experience. His current lines of research are: Methodologies for the design and evaluation of user experience, Cognitive user modeling, Interactive spaces and non-formal education. 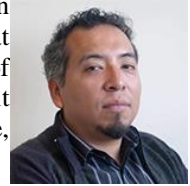